\documentclass[showpacs,amsmath,amssymb,twocolumn]{revtex4}

\begin{document}

\title{Classical and quantum Big Brake cosmology for scalar field and tachyonic models}
\author{Alexander Y. Kamenshchik}
\affiliation{Dipartimento di Fisica and INFN, Via Irnerio 46, 40126 Bologna,
Italy\\
L.D. Landau Institute for Theoretical Physics of the Russian
Academy of Sciences, Kosygin str. 2, 119334 Moscow, Russia}
\author{Serena Manti}
\affiliation{Dipartimento di Fisica,Via Irnerio 46, 40126 Bologna,
Italy\\
Scuola Normale Superiore, Piazza dei Cavalieri 7, 56126 Pisa, Italy}

\begin{abstract}
We  study a  relation between the cosmological singularities in classical and quantum theory, comparing the classical and quantum dynamics in some models possessing the Big Brake singularity - the model based on a scalar field and two models based on a tachyon-pseudo-tachyon field .
It is shown that the effect of quantum avoidance is absent for the soft singularities of the Big Brake type while 
it is present for the Big Bang and Big Crunch singularities. Thus, there is some kind of a classical - quantum 
correspondence, because  soft singularities are traversable in classical cosmology, while the strong Big Bang and Big Crunch 
singularities are not traversable. 
\end{abstract}
\pacs{98.80.Qc, 98.80.Jk, 04.60.Ds}
\maketitle

\section{Introduction}

The problem of cosmological singularities has been attracting the attention of theoreticians since the early fifties
\cite{Land,Misn-Torn,Hawk-Ell}. In sixties the general theorems about the conditions for the appearance of singularities were
proven  \cite{Hawk,Pen} and the oscillatory  regime of the approaching to the singularity \cite{BKL} called
also Mixmaster universe \cite{Misner} was discovered. The introduction of the notion of the quantum state of the universe, satisfying the Wheeler-DeWitt equation \cite{DeWitt} has stimulated the diffusion of the hypothesis that in the framework
of quantum cosmology the singularities can disappear in some sense. Namely, the probability of finding of the universe
with the parameters, which correspond to a classical cosmological singularity can be equal to zero (for a recent review see
\cite{Kiefer}).

Basically, until the end of nineties almost all the discussions about classical and quantum cosmology of singularities were devoted to the Big Bang and Big Crunch singularities, which are characterized by the vanishing value of the cosmological radius. The situation was changed after the
discovery of the phenomenon of the cosmic acceleration \cite{Riess, Perlmutter}. Such a discovery was the starting point for the formulation of cosmological models containing a special type of substance, the so-called dark energy, which, for its specific properties, was considered responsible for the accelerated expansion of the universe \cite{Sahni, Padmanabhan1, Peebles, Sami}, and consequently stimulated the study of the various possible candidates for the role of this substance.
The fundamental feature of the dark energy which produces an accelerated expansion is that it possesses a pressure $p$ such that the strong energy condition $\rho +3p >0$ is violated (here $\rho$ is the energy density). 
The construction of different cosmological models, describing dark energy,
has attracted the attention of researchers to the fact that other types of cosmological singularities do exist. First of all,
one should mention the Big Rip singularity \cite{Rip-Star,Rip} arising in the models where the phantom dark energy \cite{phantom} is present. Under phantom dark energy one understands the substance whose pressure is negative and has an
absolute value bigger than its energy density. Such a singularity is characterized by infinite values of the cosmological radius (scale factor), of its time derivative, of the Hubble parameter and its time derivative and, hence, of its energy density and pressure.

Another class of singularities includes the so-called soft or sudden singularities \cite{Barrow-old,Shtanov, Barrow, Gorini}.
They occur at a finite value of the scale factor and of its time derivative, and hence of the Hubble parameter and of the energy density, while the second derivative of the scale factor, the first derivative of the Hubble parameter and the pressure are divergent. The particularity of the Big Brake cosmological singularity, belonging to this class, consists in the fact that the time derivative
of the scale factor is equal exactly to zero. That makes this singularity especially convenient for study.
The Big Brake singularity was first described in paper \cite{Gorini}, where it has arisen in the context of a particular
cosmological model with a tachyon field, whose potential depended on the trigonometrical functions. In the same paper it was noticed that a very simple cosmological model, based on the anti-Chaplygin gas, leads unavoidably to the Big Brake singularity.
The anti-Chaplygin gas, with the equation of state $p=\frac{A}{\rho }$, where  $A$ is a positive constant, arises in the theory of wiggly strings \cite{Carter} and has got this name \cite{Gorini} in analogy with the Chaplygin gas, which satisfies an equation of state of the type $p=-\frac{A}{\rho }$ and has acquired a certain importance in cosmology as a candidate for the unification between dark energy and dark matter \cite{Kamenshchik2, Fabris}.

Starting from the anti-Chaplygin gas cosmological model and using the technique of reconstruction of the potentials for the scalar field models, one can construct the scalar field model reproducing the cosmological evolution occurring in
the anti-Chaplygin gas cosmological model. Such a potential was constructed and studied in \cite{Kamenshchik1}.
In the same paper the quantum cosmology of the corresponding model was studied and it was shown that the requirement of the normalizability of the quantum state of the universe, satisfying the Wheeler-DeWitt equation, implies the disappearance of
this quantum state at the Big Brake singularity. Thus, this result looks as confirming the hypothesis that in the framework of quantum cosmology the singularities can disappear. 
Similar researches, devoted to the properties of the solutions of the Wheeler-DeWitt equations for different cosmological models, connected 
in some way with dark energy hypothesis, have given analogous results \cite{quantum}. 
However, two question arise: first, how general is this phenomenon? In other words, should the wave function of the universe satisfying the Wheeler-DeWitt equation disappear at the values of its arguments, which classically correspond to the soft singularities?  The second question is more subtle: does the disappearance of the wave function of the universe at some values of its arguments mean that 
the relevant probability distribution disappears too? The point is that the wave function of the universe satisfying 
the Wheeler-DeWitt equation does not have a direct probabilistic interpretation \cite{barv}. To provide such an interpretation one 
has to choose a time depending gauge-fixing condition and after that one should undertake the reduction of the set of variables to the smaller 
set of the physical degrees of freedom \cite{barv}. The explicit realization of this procedure is rather complicated, but 
already study of its general features can give some interesting results.

To try to answer the questions formulated above we shall give a comparative analysis of three cosmological models, encountering the Big Brake singularity:
the scalar field model \cite{Kamenshchik1} and two  tachyon models \cite{Gorini}. The classical dynamics of the approach to
the Big Brake singularity in the tachyon model with trigonometric potential was considered in detail in paper \cite{Zoltan} while the mechanism of the crossing of this singularity was suggested in \cite{Zoltan1}. There it was shown that there was a large class of cosmological evolutions
crossing the Big Brake singularity. Thus, we have some kind of complementarity - the classical dynamics of the tachyon model with trigonometric potential \cite{Gorini}  encountering  the Big Brake singularity was studied in detail in \cite{Gorini, Zoltan,Zoltan1} and it was shown that this singularity is traversable. On the other hand the quantum dynamics of the model with the scalar field was studied
in \cite{Kamenshchik1} and there it was shown that the wave function of the universe disappears encountering such a singularity. In the present paper we shall try to fill the ``holes''
of the preceding considerations. Namely, we shall study in detail the classical dynamics of the model with the scalar field and
we shall show that the cosmological trajectory arriving to the Big Brake is unique. However, all other trajectories cross the 
soft singularities of more general kind - namely, the singularities where the deceleration is infinite, but the Hubble parameter is finite and differs from zero. The wave function of the universe disappears at such a singularity, but we shall argue that the corresponding probability is different from zero, due to the Faddeev-Popov factor arising in the process of reducing to the 
physical degrees of freedom. The situation with a 
quantum dynamics of the 
simple pseudo-tachyon model with a constant potential is quite similar. The analysis of the rather complicated tachyon model 
with trigonometric potential shows that in this case the wave function of the universe is not obliged to disappear 
at the values of its arguments, corresponding to the classical Big Brake singularity.

On the other hand, in the analysis of the Big Bang and the Big Crunch singularities, which in these models are not traversable
classically, shows that the effect of quantum avoidance is present. Thus, we have some kind of the quantum-classical correspondence 
here. For the singularities which are not traversable in classical cosmology the effect of quantum avoidance of 
singularities is present, while for the soft traversable singularities such an effect is absent.   

Concluding the Introduction we would like to remind that while the soft (sudden) singularities were known long before 
the discovery of the cosmic acceleration \cite{Barrow-old}, the development of the dark energy models has stimulated 
the study of this kind of singularities. It was shown that the presence of such singularities in some models of dark energy
does not contradict observational data on Supernovae of type Ia \cite{Zoltan, Mariusz}. However, in this paper we are not concerned with the dark energy problem. We rather study some general questions arising in the classical and quantum 
cosmology in the presence of singularities.

The structure of the paper is the following: in the second section we briefly illustrate the Big Brake cosmology,
using the model with an anti-Chaplygin gas, besides we introduce the related scalar field cosmological model; in section III we
study in detail the classical dynamics of this model; in section IV we remind the results of paper \cite{Kamenshchik1}
concerning the quantum cosmology of the model with the scalar field and compare the classical and quantum cosmology of this model; in section V we
briefly recapitulate the basic information about the tachyon cosmological model \cite{Gorini} and about its classical
dynamics \cite{Zoltan,Zoltan1} while in sec. VI we discuss its quantum dynamics. The last section is devoted to the concluding remarks.

\section{The cosmological model with the anti-Chaplygin gas and the related scalar field potential}
Let us consider a flat Friedmann-Lema\^itre-Robertson-Walker  universe
with the metric
\begin{equation}
ds^2 = dt^2-a^2(t)dl^2,
\label{Fried}
\end{equation}
filled with an anti-Chaplygin gas \cite{Carter,Gorini} with the equation of state
\begin{equation}
p=\frac{A}{\rho},
\label{anti-Chap}
\end{equation}
where $A$ is a positive constant.
The Friedmann equation is
\begin{equation}
H^2 = \rho,
\label{Fried1}
\end{equation}
where the Hubble parameter $H$ is as usual
\begin{equation}
H \equiv \frac{\dot{a}}{a},
\label{Hubble}
\end{equation}
where ``dot'' means the derivative with respect to the cosmic time $t$.
The energy conservation condition is
\begin{equation}
\dot{\rho }=-3H(\rho +p).
\label{conserv}
\end{equation}

Eqs. (\ref{anti-Chap}) and (\ref{conserv}) give immediately the dependence of the energy density $\rho$ on the cosmological radius $a$:
\begin{equation}
\rho(a)=\sqrt{\frac{B}{a^6}-A} \hspace{0.1cm},
\label{rho}
\end{equation}
where  $B>0$ is an integration constant.

Substituting the expression (\ref{rho}) into the Friedmann equation (\ref{Fried1}) one can
explicitly find the dependence of the cosmic time $t$ on $a$ (see \cite{Kamenshchik1}),
but we shall not need it. Let us notice instead that at the beginning of the cosmological
evolution, when the cosmological radius is very small we have $\rho \sim 1/a^3$ and the fluid behaves like dust. Then, when the cosmological radius tends to the critical value
$a_{*} = (B/A)^{\frac{1}{6}}$, the energy density disappears and the pressure, according to
Eq. (\ref{anti-Chap}), grows indefinitely. Thus, we encounter some kind of cosmological singularity. Let us study it in some detail.

In the vicinity of the singularity the cosmological radius can be represented as
\begin{equation}
a(t) = a_* -\Delta a(t),
\label{Delta}
\end{equation}
where $\Delta a(t)$ is a small magnitude. Substituting (\ref{Delta}) into the Friedmann
equation (\ref{Fried1}) with the expression (\ref{rho}) in the right-hand side and integrating it
in the vicinity of the time moment $t_B$ such that $a(t_B) = a_*$, we obtain
\begin{equation}
\Delta a(t)=C(t_B-t)^{4/3}\hspace{0.1cm},
\label{Delta1}
\end{equation}
where $C = 3^{\frac53}2^{-\frac73}(AB)^{\frac16}$.
Then,
\begin{eqnarray}
&&a(t)=a_*-C(t_B-t)^{4/3},\nonumber \\
&&\dot{a}(t)=\frac{4C}{3}(t_B-t)^{1/3},\nonumber \\
&&\ddot{a}(t)=-\frac{4C}{9}(t_B-t)^{-2/3}.
\label{a-t}
\end{eqnarray}
From these expressions we see that when the time tends to $t_B$, the cosmological radius
tends to the finite value $a_*$, its time derivative disappears, i.e. the expansion of the universe stops, but the second time derivative instead tends to $-\infty$, i.e. the deceleration
is infinite. Thus, the stop of expansion occurs in a singular way that justifies the name of
this singularity - Big Brake \cite{Gorini}.

Let us notice that the Big Brake singularity just like other soft singularities
possesses the important property that the Christoffel symbols at the singularity are finite
(or even zero) \cite{Laz}. Thus, the matter can pass through this singularity and then
the geometry of the spacetime can reappear \cite{Zoltan1}.  
Different aspects of the soft singularities crossing were considered also in papers \cite{cross}.
In the case of the universe filled with the anti-Chaplygin gas the things look particularly simple. One can easily see that
the expressions (\ref{a-t}) are well defined at $t > t_B$. Namely, after arriving at the point
of the Big Brake $a(t_B) = a_*$, which is at the same time the point of the maximal expansion of the universe, the universe begin contracting and this contraction culminates in the
encounter with a Big Crunch singularity at $a = 0$. Thus, the model with the anti-Chaplygin gas describes the evolution of the universe from the Big Bang to the Big Crunch passing through the soft Big Brake singularity at the moment of the maximal expansion of the universe.

However, the described evolution is the only evolution present in this model. If we want to have a little bit more rich model for the analysis of its classical and quantum dynamics, we can use rather a standard procedure of the reconstruction of  potentials of minimally coupled scalar fields \cite{Star-rec,Bar-rec,Gorini,Andr,Tron,Ser}, reproducing a given cosmological evolution. This procedure is based on the use of two equations:
\begin{equation}
\dot{\varphi}^2 = \rho + p,
\label{rec}
\end{equation}
\begin{equation}
V = \frac12(\rho-p).
\label{rec1}
\end{equation}
Here we shall give the result of the reconstruction procedure (for details see \cite{Kamenshchik1}):
\begin{equation}
V(\varphi) = \pm\frac{\sqrt{A}}{2}\left(\sinh 3\varphi -\frac{1}{\sinh 3\varphi}\right).
\label{poten-scal}
\end{equation}
As a matter of fact we have two possible potentials, which differs by the general sign.
We choose the sign ``plus''. Then, let us remember that the Big Brake occurs when the
energy density is equal to zero (the disappearance of the Hubble parameter) and the pressure is positive and infinite (an infinite deceleration). To achieve this condition, in the scalar field model it is necessary to require that the potential is negative and infinite. It
is easy to see from Eq. (\ref{poten-scal}) that this occurs when $\varphi \rightarrow 0$ being positive. Thus, to have the model with the Big Brake singularity we can consider the scalar field with a potential which is a little bit simpler than that from Eq. (\ref{poten-scal}), but still possesses rather a rich dynamics. Namely we shall study the scalar field with the potential
\begin{equation}
V = -\frac{V_0}{\varphi},
\label{poten1}
\end{equation}
where $V_0$ is a positive constant. The next section will be devoted to the analysis of the classical cosmology of the model with this potential.

\section{Classical dynamics of the cosmological model with a scalar field whose potential is inversely proportional to the field}
The Klein-Gordon equation for the scalar field with the potential (\ref{poten1}) is
\begin{equation}
\ddot{\varphi} + 3H\dot{\varphi}+\frac{V_{0}}{\varphi^{2}}=0
\label{KG}
\end{equation}
while the first Friedmann equation is
\begin{equation}
H^{2}=\frac{\dot{\varphi }^{2}}{2}-\frac{V_{0}}{\varphi} \hspace{0.1cm}.
\label{Fried2}
\end{equation}
We shall also need the expression for the time derivative of the Hubble parameter, which
can be easily obtained from Eqs. (\ref{KG}) and (\ref{Fried2}):
\begin{equation}
\dot{H}=-\frac{3}{2}\dot{\varphi }^{2} \hspace{0.1cm}.
\label{Hdot}
\end{equation}
Now we shall construct the complete classification of the cosmological evolutions
(trajectories) of our model, using Eqs. (\ref{KG})-(\ref{Hdot}).

First of all, let us announce briefly the main results of our analysis.
\begin{enumerate}
\item
The transitions between the positive and negative values of the scalar field are impossible.
\item
All the trajectories (cosmological evolutions) with positive values of the scalar field 
begin in the Big Bang singularity, then achieve a point of maximal expansion, then contract and 
end their evolution in the Big Crunch singularity.
\item   
All the trajectories with positive values of the scalar field pass through the point where the value of the scalar field 
is equal to zero. After that the value of the scalar field begin growing. The point $\varphi = 0$ corresponds to a crossing of the soft singularity.
\item 
If the moment when the universe achieves the point of the maximal expansion coincides with the moment of the  crossing of the soft singularity  then the singularity is the Big Brake. 
\item
The evolutions with the negative values of the scalar field belong to two classes - first, an infinite expansion beginning from 
the Big Bang 
and second, the evolutions obtained by the time reversion of those of the first class, which are contracting and end 
in the Big Crunch singularity. 
\end{enumerate}

To prove these results, we begin with the consideration of the universe in the vicinity of the point $\varphi = 0$.
We shall look for the leading term of the field $\varphi$ approaching this point in the form
\begin{equation}
\varphi(t) = \varphi_1(t_S-t)^{\alpha},
\label{lead}
\end{equation}
where $\varphi_1$ and $\alpha$ are positive constants and $t_S$ is the moment of the soft singularity crossing.
The time derivative of the scalar field is now 
\begin{equation}
\dot{\varphi}(t) = \alpha\varphi_1(t_S-t)^{\alpha-1}.
\label{lead1}
\end{equation}
Because of the negativity of the potential (\ref{poten1}) at positive values of $\varphi$, the kinetic term should be stronger than the potential one to satisfy the Friedmann equation (\ref{Fried2}). That implies that $\alpha 
\leq \frac23$.  However, if $\alpha < \frac23$ we can neglect the potential term and remain with the massless scalar field.
It is easy to show considering the Friedmann (\ref{Fried2}) and Klein-Gordon (\ref{KG}) equations that in this case the scalar 
field behaves like $\varphi \sim \ln (t_S-t)$, which is incompatible with the hypothesis of its smallness (\ref{lead}). Thus, 
one remains with the only choice 
\begin{equation}
\alpha = \frac23.
\end{equation}
Then, if the coefficient at the leading term in the kinetic energy is greater than that in the potential, it follows from the Friedmann
equation (\ref{Fried2})  that the Hubble parameter behaves as $(t_S-t)^{-\frac13}$ which is incompatible with Eq. (\ref{Hdot}). Thus, the leading terms of the potential and kinetic energy should cancel each other: 
\begin{equation}
\frac12\alpha^2\varphi_1^2(t_S-t)^{2\alpha-2} =\frac{V_0}{\varphi_1}(t_S-t)^{-\alpha},
\label{compare}
\end{equation}
that for $\alpha = \frac23$ gives 
\begin{equation}
\varphi_1 =\left(\frac{9V_0}{2}\right)^{\frac13}.
\label{varphi1}
\end{equation}
Hence the leading term for the scalar field in the presence of the soft singularity is 
\begin{equation}
\varphi(t) = \left(\frac{9V_0}{2}\right)^{\frac13}(t_S-t)^{\frac23}.
\label{brake}
\end{equation}
Now, integrating Eq. (\ref{Hdot}) we obtain
\begin{equation}
H(t) = 2\left(\frac{9V_0}{2}\right)^{\frac23}(t_S-t)^{\frac13} +H_S,
\label{Hbrake}
\end{equation}
where $H_S$ is an integration constant giving the value of the Hubble parameter at the moment of the soft singularity crossing.
If this constant is equal to zero, $H_S = 0$, the moment of the maximal expansion of the universe coincides with 
that of the soft singularity crossing and the universe encounters the Big Brake singularity. If $H_S \neq 0$ we have a more 
general type of the soft cosmological singularity where the energy density of the matter in the universe is different from zero.
The sign of $H_S$ can be both, positive or negative, hence, universe can pass through this singularity in the phase of its 
expansion or of its contraction.

The form of the leading term for the scalar field in the vicinity of the moment when $\varphi = 0$ (\ref{brake}) shows that, after passing the zero value, the scalar field begin growing being positive. Thus, it proves the first result from the list
presented above about impossibility of the change of the sign of the scalar field in our model. 

We have already noted that the time derivative of the scalar field had changed the sign crossing the soft singularity.
It cannot change the sign in a non-singular way because the conditions $\dot{\varphi}(t_0) = 0, \varphi(t_0) \neq 0$ are incompatible with the Friedmann  equation (\ref{Fried2}). It is seen from Eq. (\ref{brake}) that before the crossing of the soft singularity  
the time derivative of the scalar field is negative and after its crossing it is positive. The impossibility of the changing 
the sign of the time derivative of the scalar field without the soft singularity crossing implies the inevitability of the approaching of the universe to this soft singularity. Thus, the third result from the list above is proven. 

It is easy to see from Eq. (\ref{Hdot}) that the value of the Hubble parameter is decreasing during all the evolution. 
At the same time, the absolute value of its time derivative (proportional to the time derivative squared of the scalar field) 
is growing after the soft singularity crossing. That means that at some moment the Hubble parameter should change its sign 
becoming negative. The change of the sign of the Hubble parameter is nothing but the passing through the point of the maximal 
expansion of the universe, after which it begin contraction culminating in the encounter with the Big Crunch singularity.
Thus the second result from the list presented above is proven. 

Summing up, we can say that all the cosmological evolutions where the scalar field has positive values have the following structure: they begin in the Big Bang singularity with an infinite positive value of the scalar field and an infinite negative value of its time derivative, then they pass through the soft singularity where the value of the scalar field is equal to zero and where the derivative of the scalar field changes its sign. All the trajectories also pass through the point of the maximal expansion, and this passage trough the point of the maximal expansion can precede or follow the passage trough the soft singularity:
in the case when these two moments coincide ($H_S = 0$) we have the Big Brake singularity (see the result 4 from the list above). Thus, all the evolutions pass through the soft singularity, but only for one of them this singularity has a character of the Big Brake singularity. The family of the trajectories can be parameterized by the value of the Hubble parameter $H_S$ at the moment of the crossing of the soft singularity. There is also another natural parameterization of this family - we can 
characterize a trajectory by the value of the scalar field $\varphi$ at the moment of the maximal expansion of the universe and by the sign of its time derivative at this moment (if the time derivative of the scalar field is negative that means that the 
passing through the point of maximal expansion precedes the passing through the soft singularity and if the sign of this time derivative
is positive, then passage trough the point of  maximal expansion follows the passage through the soft singularity). If at the moment when the universe achieves the point of maximal expansion the value of the scalar field is equal to zero, then it is the exceptional trajectory crossing the Big Brake singularity.

For completeness, we shall say some words about the result 5, concerning the trajectories with the negative values of the scalar field. Now, both the terms in the right-hand side of the Friedmann equation (\ref{Fried2}), potential and kinetic, are positive 
and, hence, the Hubble parameter cannot disappear or change its sign. It can only tends to zero asymptotically while 
both these terms tend asymptotically to zero. Thus, in this case there are two possible regimes: an infinite expansion which begins with the Big Bang singularity and an infinite contraction which culminates in the 
encounter with the Big Crunch singularity. The second regime can be obtained by the time reversal of the first one and vice versa. Let us consider the expansion regime. 
It is easy to check that the scalar field being negative cannot achieve the zero value, because the suggestion 
$\varphi(t) = -\varphi_1(t_0-t)^{\alpha}$, where $\varphi_1 < 0, \alpha > 0$ is incompatible with the equations 
(\ref{Fried2}) and (\ref{Hdot}). Hence,  the potential term is always non-singular and at the birth of the universe from the 
Big Bang singularity the kinetic term dominates and the dynamics is that of the theory with the massless scalar field.
Namely 
\begin{equation}
\varphi(t) = \varphi_0 +\sqrt{\frac29}\ln t,\ \ H(t) = \frac{1}{3t},
\label{neg0}
\end{equation}
where $\varphi_0$ is a constant. At the end of the evolution the Hubble parameter tends to zero, while the time grows indefinitely. That means that both the kinetic and potential terms in the right-hand side of Eq. (\ref{Fried2}) should 
tend to zero. It is possible if the scalar field tends to infinity while its time derivative tends to zero.
The joint analysis of Eqs. (\ref{Fried2}) and (\ref{Hdot}) gives the following results for the asymptotic behavior 
of the scalar field and the Hubble parameter:
\begin{equation}
\varphi(t) =\tilde{\varphi}_0-\left(\frac56\right)^{\frac25}V_0^{\frac15}t^{\frac25},\ \ H(t) = \left(\frac65\right)^{\frac15}V_0^{\frac25}t^{-\frac15},
\label{neg1}
\end{equation}
where $\tilde{\varphi}_0$ is a constant.

\section{The quantum dynamics of the cosmological model with a scalar field whose potential is inversely proportional to the field}
In this section we recapitulate briefly the results of paper \cite{Kamenshchik1}, where the quantum cosmology of the model 
with a scalar field, whose potential is inversely proportional to the field, was studied. We shall try to re-interprete some of the results obtianed in  \cite{Kamenshchik1} putting them in more wide context.

As usual, we shall use the canonical formalism and the Wheeler-DeWitt equation \cite{DeWitt}. For this purpose, instead of the Friedmann metric (\ref{Fried}), we shall consider a more general metric, 
\begin{equation}
ds^2 = N^2(t)dt^2 -a^2(t)dl^2,
\label{lapse}
\end{equation}
where $N$ is the so-called lapse function. The action of the Friedmann flat model with the minimally coupled scalar field looks now as 
\begin{equation}
S = \int dt \left(\frac{a^3\dot{\varphi}^2}{2N}-a^3V(\varphi) -\frac{a\dot{a}^2}{N}\right).
\label{action}
\end{equation}
Variating the action (\ref{action}) with respect to $N$ and putting then $N=1$ we come to the standard Friedmann equation. 
Now, introducing the canonical formalism, we define the canonically conjugated momenta as 
\begin{equation}
p_{\varphi} = \frac{a^3\dot{\varphi}}{N}
\label{mom-phi}
\end{equation}
and 
\begin{equation}
p_{a} = -\frac{a\dot{a}}{N}.
\label{mom-a}
\end{equation}
The Hamiltonian is
\begin{equation}
{\cal H} = N\left(-\frac{p_a^2}{4a}+\frac{p_{\varphi}^2}{2a^3}+Va^3\right)
\label{Hamilton}
\end{equation}
and is proportional to the lapse function. The variation of the action with respect to $N$ gives the constraint  
\begin{equation}
-\frac{p_a^2}{4a}+\frac{p_{\varphi}^2}{2a^3}+Va^3 = 0,
\label{constraint}
\end{equation}
and the implementation of the Dirac quantization procedure gives the Wheeler-DeWitt equation
\begin{equation}
\left(-\frac{\hat{p}_a^2}{4a}+\frac{\hat{p}_{\varphi}^2}{2a^3}+Va^3\right)\psi(a,\varphi) = 0.
\label{WDW}
\end{equation}
Here $\psi(a,\phi)$ is the wave function of the universe and the hats over the momenta mean that the functions are substituted by operators. Introducing the differential operators representing the momenta as
\begin{equation}
\hat p_a \equiv \frac{\partial}{i \partial a},\ \ \hat p_{\varphi} \equiv \frac{\partial}{i\partial \varphi}  
\label{dif}
\end{equation}
and multiplying Eq. (\ref{WDW}) by $a^3$ we obtain the following partial differential equation:
\begin{equation}
\left(\frac{a^2}{4}\frac{\partial^2}{\partial a^2} - \frac12\frac{\partial^2}{\partial \varphi^2} +a^6V\right)\psi(a,\varphi) = 0.
\label{WDW1}
\end{equation}
Finally, for our potential inversely proportional to the scalar field we have
 \begin{equation}
\left(\frac{a^2}{4}\frac{\partial^2}{\partial a^2} - \frac12\frac{\partial^2}{\partial \varphi^2} -\frac{a^6V_0}{\varphi}\right)\psi(a,\varphi) = 0.
\label{WDW2}
\end{equation}
Note that in the equation (\ref{WDW}) and in the subsequent equations we have ignored rather a complicated problem of the choice of the ordering of  noncommuting operators, because the specification of such a choice is not essential for our analysis. 
Moreover, the interpretation of the wave function of the universe is rather an involved question \cite{Venturi,barv,Bar-Kam}. The point is that to choose the measure in the space of the corresponding Hilbert space we should fix a particular gauge condition, eliminating in such a 
way the redundant gauge degrees of freedom and introducing  a temporal dynamics into the model \cite{barv}. We shall not dwell here  on this procedure, assuming generally that 
the cosmological radius $a$ is in some way connected with the chosen time parameter and that the unique physical variable is 
 the scalar field $\varphi$. Then, it is convenient to represent the solution of Eq. (\ref{WDW2}) in the form 
\begin{equation}
\psi(a,\varphi) = \sum_{n=0}^{\infty}C_n(a)\chi_n(a,\varphi),
\label{wavefunction}
\end{equation}
where the functions $\chi_n$ satisfy the equation
\begin{equation}
\left(- \frac12\frac{\partial^2}{\partial \varphi^2} -\frac{a^6V_0}{\varphi}\right)\chi(a,\varphi) = -E_n(a)\chi_n(a,\varphi),
\label{chi}
\end{equation}
while the functions $C_n(a)$ satisfy the equation 
\begin{equation}
\frac{a^2}{4}\frac{\partial^2 C_n(a)}{\partial a^2} = E_n(a)C_n(a),
\label{C}
\end{equation}
where $n=0,1,\ldots$.
Requiring the normalizability of the functions $\chi_n$ on the interval $0 \leq \varphi < \infty$, which, in turn, implies   their non-singular behavior at $\varphi =0$ and $\varphi \rightarrow \infty$, and using the considerations similar to those used in the analysis of the Schr\"odinger equation for the hydrogen-like atoms, one can show that the acceptable values of the functions $E_n$ are 
\begin{equation}
E_n = \frac{V_0a^{12}}{2(n+1)^2},
\label{E}
\end{equation}
while the corresponding eigenfunctions are 
\begin{equation}
\chi_n(a,\varphi) = \varphi \exp\left(-\frac{V_0a^{6}\varphi}{n+1}\right)L_n^1\left(\frac{2V_0a^6\varphi}{n+1}\right),
\label{Laguerre}
\end{equation}
where $L_n^1$ are the associated Laguerre polynomials. 

Rather often the fact that the wave function of the universe disappears at the values of the cosmological parameters
corresponding to some classical singularity is interpreted as an avoidance of such singularity. 
However, in the case of the soft singularity considered in the model at hand, such an interpretation 
does not look too convincing. 
Indeed, one can have a temptation to think that the probability of finding of the universe in the soft singularity state
characterized by the vanishing value of the scalar field is vanishing because the expression for functions  (\ref{Laguerre}) entering into the 
expression for the wave function of the universe (\ref{wavefunction}) is proportional to $\varphi$. However, the wave function 
(\ref{wavefunction}) can hardly have a direct probabilistic interpretation. Instead, one should choose some reasonable time-dependent gauge, identifying some combination of  variables with an effective time parameter, and interpreting other 
variables as physical degrees of freedom \cite{barv}. The definition of the wave function of the universe in terms of these 
physical degrees of freedom is rather an involved question; however, we are in a position to make some semi-qualitative 
considerations. The  reduction of the initial set of variables to the smaller set of physical degrees of freedom implies the appearance of the Faddeev-Popov determinant which as usual is equal to the Poisson bracket of the gauge-fixing condition 
and the constraint. Let us, for example, choose as a gauge-fixing condition the identification of the new ``physical'' time 
parameter with the Hubble parameter $H$ taken with the negative sign. Such an identification is reasonable, because as it follows from Eq. (\ref{Hdot}) 
the variable $H(t)$ is monotonously decreasing. The volume $a^3$ is the variable canonically conjugated to the Hubble variable. Thus, the Poisson bracket between the gauge-fixing condition $\chi = H - T_{\rm phys}$ and the constraint (\ref{constraint}) includes 
the term proportional to the potential of the scalar field, which is inversely proportional to this field itself. Thus, the singularity in $\varphi$ arising in the Faddeev-Popov determinant can cancel zero, arising in ({\ref{Laguerre}).

 Let us confront this situation with that of the Big Bang and Big Crunch singularities. As it was seen in Sec. III 
such singularities classically arise at infinite values of the scalar field. To provide the normalizability of the wave function  one should have the integral on the values of the scalar field $\varphi$ convergent, when $|\varphi| \rightarrow \infty$. 
That means that, independently of details connected with the gauge choice, not only the wave function of the universe but also  the probability density of scalar field values  should decrease 
rather rapidly when the absolute value of the scalar field is increasing. Thus, in this case, the effect of the quantum avoidance of the classical singularity is present.

\section{The tachyon cosmological  model with the trigonometric potential}
The tachyon field, born in the context of the string theory \cite{Sen},
provides an example of matter having a large enough negative pressure to produce an acceleration of the expansion rate of the universe. Such a field is today considered as one of the possible candidates for the role of dark energy and, also for this reason, in the recent years it has been intensively studied. The tachyon models represent a subclass of the models with non-standard kinetic terms \cite{k-ess}, which descend from the Born-Infeld model, invented already in thirties \cite{Born}.
Before considering the model with the trigonometric potential \cite{Gorini}, possessing the Big Brake singularity, we write down the general formulae of the tachyon cosmology. 

The Lagrangian of the tachyon field $T$ is 
\begin{equation}
L = -\sqrt{-g}V(T)\sqrt{1-g^{\mu\nu}T_{,\mu}T_{,\nu}}
\label{tach}
\end{equation}
or, for the spatially homogeneous tachyon field, 
\begin{equation}
L =-\sqrt{-g}V(T)\sqrt{1-\dot{T}^2},
\label{tach1}
\end{equation}
where $g$ is the determinant of the metric.
The energy density and the pressure of this field are respectively 
\begin{equation}
\rho = \frac{V(T)}{\sqrt{1-\dot{T}^2}}
\label{tach2}
\end{equation}   
and
\begin{equation}
p = -V(T)\sqrt{1-\dot{T}^2},
\label{tach3}
\end{equation} 
while the field equation is 
\begin{equation}
\frac{\ddot{T}}{1-\dot{T}^2} + 3H\dot{T} +\frac{V_{,T}}{V(T)} = 0.
\label{KGT}
\end{equation}

We shall introduce also the pseudo-tachyon field with the Lagrangian \cite{Gorini} 
\begin{equation}
L = \sqrt{-g}W(T)\sqrt{\dot{T}^2-1}
\label{pseud}
\end{equation}
and with the energy density 
\begin{equation}
\rho = \frac{W(T)}{\sqrt{\dot{T}^2-1}}
\label{pseud1} 
\end{equation}
and the pressure
\begin{equation}
p = W(T)\sqrt{\dot{T}^2-1}.
\label{pseud2}
\end{equation}
The Klein-Gordon equation for the pseudo-tachyon field is 
\begin{equation}
\frac{\ddot{T}}{1-\dot{T}^2} + 3H\dot{T} +\frac{W_{,T}}{W(T)} = 0.
\label{KGTp}
\end{equation}

We shall also write down the equations for the time derivative of the Hubble parameter in the tachyon and pseudo-tachyons models:
\begin{equation}
\dot{H} = -\frac32\frac{V(T)\dot{T}^2}{\sqrt{1-\dot{T}^2}},
\label{Hdott}  
\end{equation}
\begin{equation}
\dot{H} = -\frac32\frac{W(T)\dot{T}^2}{\sqrt{\dot{T}^2-1}}.
\label{Hdotpt}  
\end{equation}
We see that the Hubble parameter in both these models is decreasing just like in the scalar field model (see Eq. (\ref{Hdot})).

Note that for the case when the potential of the tachyon field $V(T)$ is a constant, 
the cosmological model with this tachyon coincides with the cosmological model with the 
Chaplygin gas \cite{FKS}.  Analogously, the pseudo-tachyon model with the constant potential 
coincides with the model with the anti-Chaplygin gas. Hence, its classical dynamics is 
that described in the section II. However, it is of interest for us to integrate explicitly 
the Klein-Gordon equation (\ref{KGTp}) for the pseudo-tachyon field with the constant potential. The result is 
\begin{equation}
\dot{T}^2 = \frac{1}{1-\frac{a^6}{a_*^6}},
\label{pseud3}
\end{equation}    
where the initial conditions for the evolution are fixed by the choice of the radius $a_*$ 
at which the universe encounters the Big Brake. Note that, when the universe tends to the Big Brake, the time derivative of the pseudo-tachyon field tends to $\pm \infty$ and while the universe 
encounters the Big Bang and Big Crunch singularities the time derivative of $T$ tends to 
$\pm 1$. In both the cases the behavior of the time derivative of the pseudo-tachyon field
going to the singularities does not depend on the particular trajectory, parametrized by 
the value $a_*$.

Now we shall study  a very particular tachyon potential depending on the trigonometrical functions which was 
suggested in the paper \cite{Gorini}. Its form is 
\begin{eqnarray}
&&V(T) = \frac{\Lambda}{\sin^2\frac32\sqrt{\Lambda(1+k)}T}\nonumber \\
&&\times \sqrt{1-(1+k)\cos^2\frac32\sqrt{\Lambda(1+k)}T},
\label{poten}
\end{eqnarray} 
where $\Lambda$ is a positive constant and $k$ is a parameter, which is chosen in the interval
$-1 < k < 1$. The case of the positive values of the parameter $k$ is  especially interesting. 
In this case one subset of the cosmological evolutions is infinite and tends to a deSitter regime of the exponential expansion with the asymptotic value of the tachyon field  
$T = T_0 =\frac{\pi}{3\sqrt{\Lambda(1+k)}}$. Other trajectories go to the points of the two-dimensional phase space $(T,\dot{T})$, where the field acquires the values 
$T=T_3=\frac{2}{3\sqrt{\Lambda(1+k)}}{\rm arccos}\frac{1}{\sqrt{1+k}}$ 
or $T=T_4=\frac{2}{3\sqrt{\Lambda(1+k)}}\left(\pi-{\rm arccos}\frac{1}{\sqrt{1+k}}\right)$
and where the expression under the sign of the square root of the potential (\ref{poten}) 
vanishes. At the same moment, the time derivative of the tachyon field becomes equal to 
$\pm 1$ and hence the other square root in the Lagrangian (\ref{tach}) vanishes too.
It was shown in paper \cite{Gorini} that after that, the transformation of the tachyon into the pseudo-tachyon becomes unavoidable. 
The potential $W(T)$ for the pseudo-tachyon is obtained from the potential $V(T)$ for the 
tachyon (\ref{poten}) by the change of the sign of the expression under the square root. 
What happens after that? Let us suppose that 
the trajectory of the tachyon field crosses the point $T = T_3, \dot{T} = -1$. Then, after crossing 
this point the universe tends to the Big Brake singularity in the regime described 
by the following formulae \cite{Zoltan1}:
\begin{equation}
T = T_{BB} + \left(\frac{4}{3W(T_{BB})}\right)^{\frac13}(t-t_{BB})^{\frac13},
\label{TBB}
\end{equation}
\begin{equation}
H = \left(\frac{9W^2(T_{BB})}{2}\right)
^{\frac13}(t-t_{BB})^{\frac13},
\label{HBB}
\end{equation}
where $t_{BB}$ is the moment when the universe encounters the Big Brake singularity
and $T_{BB}$ is the value of the tachyon field at this moment. It was shown that 
$T_{BB}$ can accept the values in the interval $0 < T_{BB} \leq T_3$ \cite{Gorini} and hence the cosmological trajectories encountering the Big Brake singularity constitute an infinite one-parameter set, whose elements can be parametrized by the value of the tachyon field $T_{BB}$. 

Then, after the Big Brake crossing the cosmological expansion is followed by a contraction,
which culminates at the encounter with the Big Crunch singularity, which occurs at 
$T = 0$ and $\dot{T} = -\sqrt{\frac{1+k}{k}}$ \cite{Zoltan1}.

\section{The quantum cosmology of the tachyon and the pseudo-tachyon field}
Now, we would like to construct the Hamiltonian formalism for the tachyon and pseudo-tachyon fields. Using the metric (\ref{lapse}), one can see that the contribution of the tachyon field into the action is 
\begin{equation}
S=-\int dt Na^3V(T)\sqrt{1-\frac{\dot{T}^2}{N^2}}.
\label{ac-Ham}
\end{equation} 
The conjugate momentum for $T$ is 
\begin{equation}
p_T = \frac{a^3V\dot{T}}{N\sqrt{1-\frac{\dot{T}^2}{N^2}}}.
\label{pT}
\end{equation}
and so the velocity can be expressed as 
\begin{equation}
\dot{T} = \frac{Np_T}{\sqrt{p_T^2+a^6V^2}}.
\label{vel-mom}
\end{equation}
The Hamiltonian of the tachyon field is now 
\begin{equation}
{\cal H} = N\sqrt{p_T^2+a^6V^2}.
\label{ham-tach}
\end{equation}

Analogously, for the pseudo-tachyon field, we have 
\begin{equation}
p_T = \frac{a^3W\dot{T}}{N\sqrt{\frac{\dot{T}^2}{N^2}-1}},
\label{ps}
\end{equation}
\begin{equation}
\dot{T} = \frac{Np_T}{\sqrt{p_T^2-a^6W^2}}
\label{ps1}
\end{equation}
and
\begin{equation}
{\cal H} = N\sqrt{p_T^2-a^6W^2}.
\label{ham-tach1}
\end{equation}
In what follows it will be convenient  for us to fix the lapse function as $N=1$. 

Now, adding the gravitational part of the Hamiltonians and quantizing the corresponding observables, we obtain the following Wheeler-DeWitt equations for the tachyons 
\begin{equation}
\left(\sqrt{\hat{p}_T^2+a^6V^2} - \frac{a^2\hat{p}_{a}^2}{4}\right)\psi(a,T) = 0
\label{WDWT}
\end{equation}
and for the pseudo-tachyons  
\begin{equation}
\left(\sqrt{\hat{p}_T^2-a^6W^2} - \frac{a^2\hat{p}_{a}^2}{4}\right)\psi(a,T) = 0.
\label{WDWTp}
\end{equation}

The study of the Wheeler-DeWitt equation for the universe filled with a tachyon or a pseudo-tachyon field is rather a difficult task because the Hamiltonian depends non-polynomially on the conjugate momentum of such fields. However, one can come to interesting  conclusions, considering some particular models. 

First of all, let us consider a model with the pseudo-tachyon field having a constant potential. 
In this case the Hamiltonian in Eq. (\ref{WDWTp}) does not depend on the field $T$. Thus, it is more convenient to use the representation of  the quantum state of the universe 
where it depends on the coordinate $a$ and the momentum $p_T$. Then the Wheeler-DeWitt equation will have the following form:
\begin{equation}
\left(\sqrt{p_T^2-a^6W^2} + \frac{a^2}{4}\frac{\partial^2}{\partial a^2}\right)\psi(a,p_T) = 0.
\label{WDWTp1}
\end{equation}
It becomes algebraic in the variable $p_T$. Now, we see that the Hamiltonian is well defined at $p_T^2 \geq a^6W^2$. Looking at the limiting value $p_T^2 = a^6W^2$ and comparing it with the relation (\ref{ps1}) we see that it corresponds to $\dot{T}^2 \rightarrow \infty$, which,
in turn, corresponds to the encounter with the Big Brake singularity as was explained in the section V. The only way to ``neutralize'' the  values of $p_T$, which imply the negativity 
of the expression under the square root in the left-hand side of Eq. (\ref{WDWTp1}), is to require that  the wave function of the universe is such that 
\begin{equation}
 \psi(a,p_T) = 0\ \ {\rm at} \ p_T^2 \leq a^6W^2.
 \label{brake-cond}
 \end{equation}
 The last condition could be considered as a hint on 
 the quantum avoidance of the Big Brake singularity. However, as it was explained in Sec. IV on the example of the scalar field 
model, to speak about the probabilities in the neighborhood of the point where the wave function of the universe vanishes,
it is necessary to realize the procedure of the reduction of the set of variables to a smaller set of physical degrees of freedom. Now, let us suppose that the gauge-fixing condition is chosen in such a way that the role of time is played by a Hubble parameter. In this case the Faddeev-Popov determinant, equal to the Poisson bracket between the gauge-fixing condition 
and the constraint, will be inversely proportional to the expression $\sqrt{\hat{p}_T^2-a^6W^2}$ (see Eq. (\ref{ham-tach1})), which tends to zero at the 
moment of the encounter with the Big Brake singularity. Thus, in the case of a pseudo-tachyon model, just like in the case 
of the cosmological model based on the scalar field, the Faddeev-Popov determinant introduces the singular factor, which 
compensates the vanishing of the wave function of the universe.

 What can we say about the Big Bang and the Big Crunch singularities in this model? 
  It was noticed in the preceding section that at these singularities $\dot{T}^2 = 1$. 
  From the relation (\ref{ps1}) it follows that such values of $\dot{T}$ correspond to 
  $|p_T| \rightarrow \infty$. A general requirement of the normalizability of the wave function of the universe implies the vanishing of $\psi(a,p_T)$ at $p_T \rightarrow \pm \infty$ which 
  signifies the quantum avoidance of the Big Bang and the Big Crunch singularities. 
It is quite natural, because these singularity are not traversable in classical cosmology. 

Now we consider the tachyon cosmological model with the trigonometric potential, whose 
classical dynamics was briefly sketched in the preceding section. In this case the Hamiltonian depends on both the tachyon field $T$ and its momentum $p_T$. The dependence of the expression under the square root on  $T$ is more complicated than that on $p_T$. Hence,  it does not make sense to use the representation $\psi(a,p_T)$ instead of $\psi(a,T)$. 
Now, we have under the square root  the second order differential operator $-\frac{\partial^2}{\partial T^2}$, which is positively defined, and the function  $-a^6W^2(T)$, which is negatively defined. The complete expression should not be negative, but what does it mean in our case? It means that we should choose such wave functions for which the quantum average of the 
operator $\hat{p}_T^2-a^6W^2(T)$ is non-negative: 
\begin{eqnarray}
&&\langle \psi |\hat{p}_T^2-a^6W^2(T)|\psi \rangle \nonumber \\ 
&&= \int {\cal D}T \psi^*(a,T) \left(-\frac{\partial^2}{\partial T^2} -a^6W(T)^2\right)\psi(a,T) \geq 0.
\label{non-neg}
\end{eqnarray}
Here the symbol ${\cal D}T$ signifies the integration on the tachyon field $T$ with some measure. 
It is easy to guess that the requirement (\ref{non-neg}) does not imply the disappearance of the wave function $\psi(a,T)$ 
at some range or at some particular values of the tachyon field, and one can always construct a wave 
function which is different from zero everywhere and thus does not show the phenomenon of the quantum avoidance of singularity. However, the forms of the potential $V(T)$ given by Eq. (\ref{poten}) and of the corresponding potential $W(T)$ for the 
pseudo-tachyon field arising in the same model \cite{Gorini} are too cumbersome to construct such functions explicitly. Thus,
to illustrate our statement, we shall consider a more simple toy model. 

Let us consider the Hamiltonian 
\begin{equation}
{\cal H} = \sqrt{\hat{p}^2 - V_0x^2},
 \label{toy}
\end{equation}
where $\hat{p}$ is the conjugate momentum of the coordinate $x$ and $V_0$ is some positive constant. Let us choose as a wave function a Gaussian function 
\begin{equation}
\psi(x) = \exp(-\alpha x^2),
\label{Gauss}
\end{equation}
where $\alpha$ is a positive number and we have omitted the normalization factor, which is not essential in the present context.
Then the condition (\ref{non-neg}) will look like 
\begin{eqnarray}
&&\int dx \exp(-\alpha x^2) \left(-\frac{d^2}{dx^2}-V_0x^2\right)\exp(-\alpha x^2) \nonumber \\
&&= \sqrt{\frac{\pi}{2}}\left(\frac34\sqrt{\alpha}-
\frac{V_0}{2\alpha^{\frac32}}\right) \geq 0,
\label{non-neg1}
\end{eqnarray}
which can be easily satisfied if 
\begin{equation}
\alpha \geq \sqrt{\frac23 V_0}.
\label{non-neg2}
\end{equation}

Thus, we have seen that for this very simple model one can always  choose such a quantum state, which does not disappear at any value of the coordinate $x$ and which guarantees the positivity of the quantum average of the operator, which is not generally 
positively defined. Coming back to  our cosmological model we can say that the requirement of the well-definiteness of the 
pseudo-tachyon part of the Hamiltonian operator in the Wheeler-DeWitt equation does not imply the disappearance of the 
wave function of the universe at some values of the variables and thus, does not reveal the effect of the quantum avoidance of the cosmological singularity.   

At the end of this section we would like also to analyze the Big Bang and Big Crunch singularities in the tachyon model with the
trigonometrical potential. As was shown in paper \cite{Gorini} the Big Bang singularity can occur in two occasions (the same is true also for the Big Crunch singularity \cite{Zoltan1}) - either $W(T) \rightarrow \infty$ (for example for $T \rightarrow 0$) 
or at $\dot{T}^2 =1, W(T) \neq 0$. One can see from Eqs. (\ref{tach2}) and (\ref{ps}) that when the universe approaches these singularities  the momentum $p_T$ tends to infinity. As was explained before, the wave function of the universe in the momentum representation should vanish at $|p_T| \rightarrow \infty$ and hence,  we have the effect of the quantum avoidance.

\section{Concluding remarks}
We have studied a relation between the cosmological singularities in classical and quantum theory, comparing the classical and quantum dynamics in some models possessing the Big Brake singularity - the model based on a scalar field and two models based on a tachyon (pseudo-tachyon) field.
It was shown that in the tachyon model with the trigonometrical potential \cite{Gorini} the wave function of the universe is not obliged to vanish in the range of the variables corresponding to the appearance of the classical Big Brake singularity. 
In a more simple pseudo-tachyon cosmological model the wave function, satisfying the Wheeler-DeWitt equation
and depending on the cosmological radius and the pseudo-tachyon  field, disappears at the Big Brake singularity. 
However, the transition to the wave function depending only on the reduced set of physical degrees of freedom implies 
the appearance of the Faddeev-Popov factor, which is singular and which singularity compensates the terms, responsible 
for the vanishing of the wave function of the universe.
Thus, in both these cases, the effect of the quantum avoidance of the Big Brake singularity is absent. 

In the case of the scalar field model with the potential inversely proportional to this field, all the classical trajectories
pass through a soft singularity (which for one particular trajectory is exactly the Big Brake). The wave function of the universe 
disappears at the vanishing value of the scalar field which classically corresponds to the soft singularity. However, also 
in this case the Faddeev-Popov factor arising at the reduction to the physical degrees of freedom provides nonzero value 
of the probability of finding of the universe at the soft singularity. 

In spite of the fact that we have considered 
some particular scalar field and tachyon-pseudo-tachyon models, our main conclusions were based on rather general properties of 
these models. Indeed, in the case of the scalar field we have used the fact that its potential at the soft singularity should be 
negative and divergent, to provide an infinite positive value of pressure. In the case of the pseudo-tachyon field both the possible vanishing of the wave function of the universe and its ``re-emergence'' in the process of reduction were connected 
with the general structure of the contribution of such a field into the super-Hamiltonian constraint (\ref{ham-tach1}).   
Note that in the case of the tachyon model with the trigonometric potential, the wave function does not disappear at all. 

On the other hand we have seen that for the Big Bang and Big Crunch singularities not only the  wave functions of the universe but also the corresponding probabilities disappear when the universe is approaching to  the corresponding values of the fields under consideration,
and this fact is also connected with rather general properties of the structure of the Lagrangians of the theories.
Thus, in these cases the effect of quantum avoidance of singularities takes place. 

One can say that there is some kind of a classical - quantum 
correspondence here. The soft singularities are traversable at the classical level (at least for simple homogeneous and isotropic Friedmann models) and the effect of quantum avoidance of singularities is absent. The strong Big Bang and Big Crunch 
singularities cannot be passed by the universe at the classical level, and the study of the Wheeler-DeWitt equation 
indicates the presence of the quantum singularity avoidance effect.  

It would be interesting also to find  examples of the absence of the effect of the quantum avoidance of singularities, 
for the singularities of the Big Bang--Big Crunch type. Note that the interest to the study of the possibility 
of crossing of such singularities is growing and some models treating this phenomenon have been elaborated during last few years \cite{Bars}.   

\section*{Acknowledgements}
We are grateful to A.O. Barvinsky and C. Kiefer for fruitful discussions and to P.V. Moniz and M. Bouhmadi-Lopez for useful 
correspondence.
This work was partially supported by the RFBR grant 11-02-00643.

\end{document}